\documentclass[twocolumn,prc,superscriptaddress,unsortedaddress,showpacs,preprintnumbers,amsmath,amssymb]{revtex4}
\usepackage{tipa}
\usepackage{amssymb}
\usepackage{graphicx}% Include figure filesph
\usepackage{dcolumn}% Align table columns on decimal point
\usepackage{bm}% bold math
\usepackage{CJK}
\usepackage{CJKulem}
\def\beq{\begin{equation}}
\def\eeq{\end{equation}}
\def\bea{\begin{eqnarray}}
\def\eea{\end{eqnarray}}

\def\fun#1#2{\lower3.6pt\vbox{\baselineskip0pt\lineskip.9pt
  \ialign{$\mathsurround=0pt#1\hfil##\hfil$\crcr#2\crcr\sim\crcr}}}
\begin{document}
\preprint{}

\title{
Correlation between $\alpha$-decay Energies of Superheavy Nuclei
Involving Effect of Symmetry Energy}
\author{Jianmin Dong}
\affiliation{Institute of Modern Physics, Chinese Academy of
Sciences, Lanzhou 730000, China} \affiliation{Graduate University of
Chinese Academy of Sciences, Beijing 100049, China}
 \affiliation{School of Nuclear
Science and Technology, Lanzhou University, Lanzhou 730000, China}
\affiliation{Institute for Theoretical Physics,
Justus-Liebig-University, D-35392 Giessen, Germany}
\author{Wei Zuo}\email[ ]{zuowei@impcas.ac.cn}
\affiliation{Institute of Modern Physics, Chinese Academy of
Sciences, Lanzhou 730000, China}
 \affiliation{School of Nuclear
Science and Technology, Lanzhou University, Lanzhou 730000, China}
\author{Werner Scheid} \affiliation{Institute
for Theoretical Physics, Justus-Liebig-University, D-35392 Giessen,
Germany}

\begin{abstract}
A formula for the relationship between the $\alpha$-decay energies
($Q$ values) of superheavy nuclei (SHN) is presented, which is
composed of the effects of Coulomb energy and symmetry energy. It
can be employed not only to validate the experimental observations
and measurements to a large extent, but also to predict the $Q$
values of heaviest SHN with a high accuracy generally which will be
very useful for future experiments. Furthermore, the shell closures
in superheavy region and the effect of the symmetry energy on the
stability of SHN against $\alpha$-decay are discussed with the help
of this formula.

\end{abstract}


\maketitle

The synthesis and identification of superheavy nuclei (SHN) have
been receiving a worldwide attention since the prediction of the
existence of superheavy island in 1960s. But where the closed shells
are located in the superheavy region is less certain, depending on
the model employed. The experimental investigations are thus crucial
and a series of experimental efforts so far have been focused on the
direct production of SHN in heavy ion fusion reactions. The
superheavy elements with $Z = 107-112$ have been successfully
produced at GSI, Darmstadt, in cold-fusion reactions \cite{GSI}.
Several new elements with $Z = 113-118$ have been discovered at
JINR-FLNR, Dubna, using hot-fusion evaporation reactions with the
neutron-rich $^{48}$Ca beam and actinide targets \cite{YY1}. The
element 114 was independently confirmed recently by the LBNL in the
USA \cite{LBNL1} and GSI \cite{GSI0}. A superheavy element isotope
$^{285}114$ was observed in LBNL last year \cite{LBNL2}, and an
isotope of $Z = 113$ has been identified at RIKEN, Japan \cite{Jap}.
Thus up to now superheavy elements with $Z = 104-118$ have been
synthesized in experiment and hence it offers the possibility to
study the heaviest known nuclear island of stability with greater
detail.

Superheavy nuclei with atomic numbers beyond 110 predominantly
undergo sequential $\alpha$-decay terminated by spontaneous fission
\cite{GSI}, leading to $\alpha$-decay that is one efficient approach
to identify new nucleus via the observation of $\alpha$-decay chain
and to extract some information about their stability. In experiment
one usually measures the $\alpha$-decay $Q$ values and half-lives,
while one of the major goals of theory is to predict the half-lives
to serve the experimental design. As one of the crucial quantity for
a quantitative prediction of decay half-life, $Q$ value strongly
affects the calculation of the half-life due to the exponential law.
Therefore, it is extremely important and necessary to obtain an
accurate theoretical $Q$ value in a reliable half-life prediction
during the experiment design. However, the existing microscopic
nuclear many-body approaches do not achieve a very good accuracy. In
this study, we propose a new approach to calculate the
$\alpha$-decay energy with a high accuracy for the superheavy
elements above 110. In our previous work \cite{dong}, a formula was
proposed for $\alpha$-decay $Q$ value of SHN based on a liquid drop
model. Taking no account of the shell energy it gives as
\begin{equation}
Q(\text{MeV})=aZA^{-4/3}(3A-Z)+b\left( \frac{N-Z}{A}\right) ^{2}+e,
\label{AA}
\end{equation}
with $a=4a_{c}/3=0.9373$, $b=-4a_{sym}=-99.3027$ and $e=-27.4530$
\cite{dong}. Here $Z$, $N$ and $A$ are the proton, neutron and mass
numbers of the parent nuclei, respectively. The first two terms on
the right hand side are the contributions of Coulomb energy and
symmetry energy, respectively. The nuclear symmetry energy plays an
important role in astrophysics \cite{star1,star3}, the structure of
exotic nuclei and the dynamics of heavy ion reactions
\cite{nuclear4,nuclear5,nuclear6}. In this Letter, the effect of
symmetry energy on the stability of SHN against $\alpha$-decay is
going to be shown.

Here we study the relationship between the $Q$ values of the
neighboring SHN taking Eq. (\ref{AA}) as the starting point but we
do not use the parameters in Ref. \cite{dong} any longer. With
$\beta =(N-Z)/A$ denoting the isospin asymmetry and $Z=A(1-\beta
)/2$, we obtain
\begin{equation}
\frac{Q_{2}-Q_{1}}{\beta _{2}-\beta _{1}}\approx \frac{\partial
Q}{\partial \beta }=-\frac{2}{3}a_{c}A^{2/3}(\beta
+2)-8a_{sym}\beta.\label{B}
\end{equation}
Once the decay energy $Q_{1}$ of a reference nucleus $^{A}Z_{1}$ is
known, the $Q_{2}$ values of the other nucleus $^{A}Z_{2}$ (target
nucleus) with the same mass number $A$ can be estimated by
\begin{equation}
Q_{2}=Q_{1}-(\beta _{2}-\beta _{1})\left[
\frac{2}{3}a_{c}A^{2/3}(\beta +2)+8a_{sym}\beta \right],\label{BB}
\end{equation}
with $\beta =(\beta _{1}+\beta _{2})/2$ and $a_{c}$=0.71. The mass
dependence of the symmetry energy coefficient is given by
Danielewicz and Lee \cite{da} as
$a_{\text{sym}}=c_{\text{sym}}(1+\kappa A^{-1/3})^{-1}$, where
$c_{\text{sym}}$ is the volume symmetry energy coefficient of the
nuclei and $\kappa$ is the ratio of the surface symmetry coefficient
to the volume symmetry coefficient. Here $c_{\text{sym}}=31.1$ and
$\kappa=2.31$ are taken from the results of Ref. \cite{WN2} without
including the uncertainty.

Apart from ($A$,$\beta$) discussed above, ($Z$,$\beta$) or
($N$,$\beta$) can be also adopted as variables. By an analogous
derivation, the correlation between the $Q$ values of the nuclei
belonging to an isotope chain with a proton number $Z$ is given by
\begin{eqnarray}
Q_{2} &=&Q_{1}-(\beta _{2}-\beta _{1})\times\nonumber  \\
&&\left[ \frac{2^{5/3}}{9}a_{c}Z^{2/3}(1-\beta )^{-2/3}(1+2\beta
)+8a_{sym}\beta \right].\label{CC}
\end{eqnarray}
and that of the nuclei belonging to an isotone chain with a neutron
number $N$ is given by
\begin{eqnarray}
Q_{2} &=&Q_{1}-(\beta _{2}-\beta _{1})\times   \notag \\
&&\left[ \frac{2^{5/3}}{9}a_{c}N^{2/3}(1+\beta )^{-5/3}(11+5\beta
+2\beta ^{2})+8a_{sym}\beta \right] .  \label{DD}
\end{eqnarray}
In general, if one selects $\xi =xZ+yN$ and $\beta$ as variables,
the relationship between the $Q$ values of $\alpha$-decay can be
written as
\begin{eqnarray}
Q_{2} &=&Q_{1}-(\beta _{2}-\beta _{1})\times
\{\frac{2^{5/3}}{9}a_{c}\xi
^{2/3}[(1-\beta )x+ \nonumber\\
&&(1+\beta )y]^{-5/3}[(1+\beta -2\beta ^{2})x+ \nonumber\\
&&(11+5\beta +2\beta ^{2})y]+8a_{sym}\beta \},\label{EE}
\end{eqnarray}
where $x$ and $y$ are integers and $\left\vert x\right\vert
^{2}+\left\vert y\right\vert ^{2}\neq 0$ with $Z=(1-\beta )\xi
/\left[ (1-\beta )x+(1+\beta )y\right] $ and $N=(1+\beta )\xi
/\left[ (1-\beta )x+(1+\beta )y\right] $. Here only the differences
of the symmetry energy effect ($a_{sym}$ term) together with the
differences of Coulomb energy effect ($a_{c}$ term) between a
reference nucleus and a target one contribute to this correlation.
The isospin dependence of the symmetry energy coefficient
$a_{\text{sym}}$ is neglected here because the $a_{\text{sym}}$
changes quite slightly between the neighboring nuclei.  With this
formula, the $Q$ values of target nuclei can be obtained by any
neighboring nuclei. Eqs. (\ref{BB}), (\ref{CC}) and (\ref{DD}) are
the special cases of Eq. (\ref{EE})--that is, $x=y=1$ for Eq.
(\ref{BB}), $x=1$, $y=0$ for Eq. (\ref{CC}) and $x=0$, $y=1$ for Eq.
(\ref{DD}), respectively.

%%%%%%%%%%%%%%%%%%%%%%%%%%%%%%%%%%%%%%%%%%%%%%%%%%%%%%%%%%%%%%%%%%%%%%%%%%%%%%%%%%  Figure 1
\begin{figure*}[htbp]
\begin{center}
\includegraphics[width=1.08\textwidth,clip]{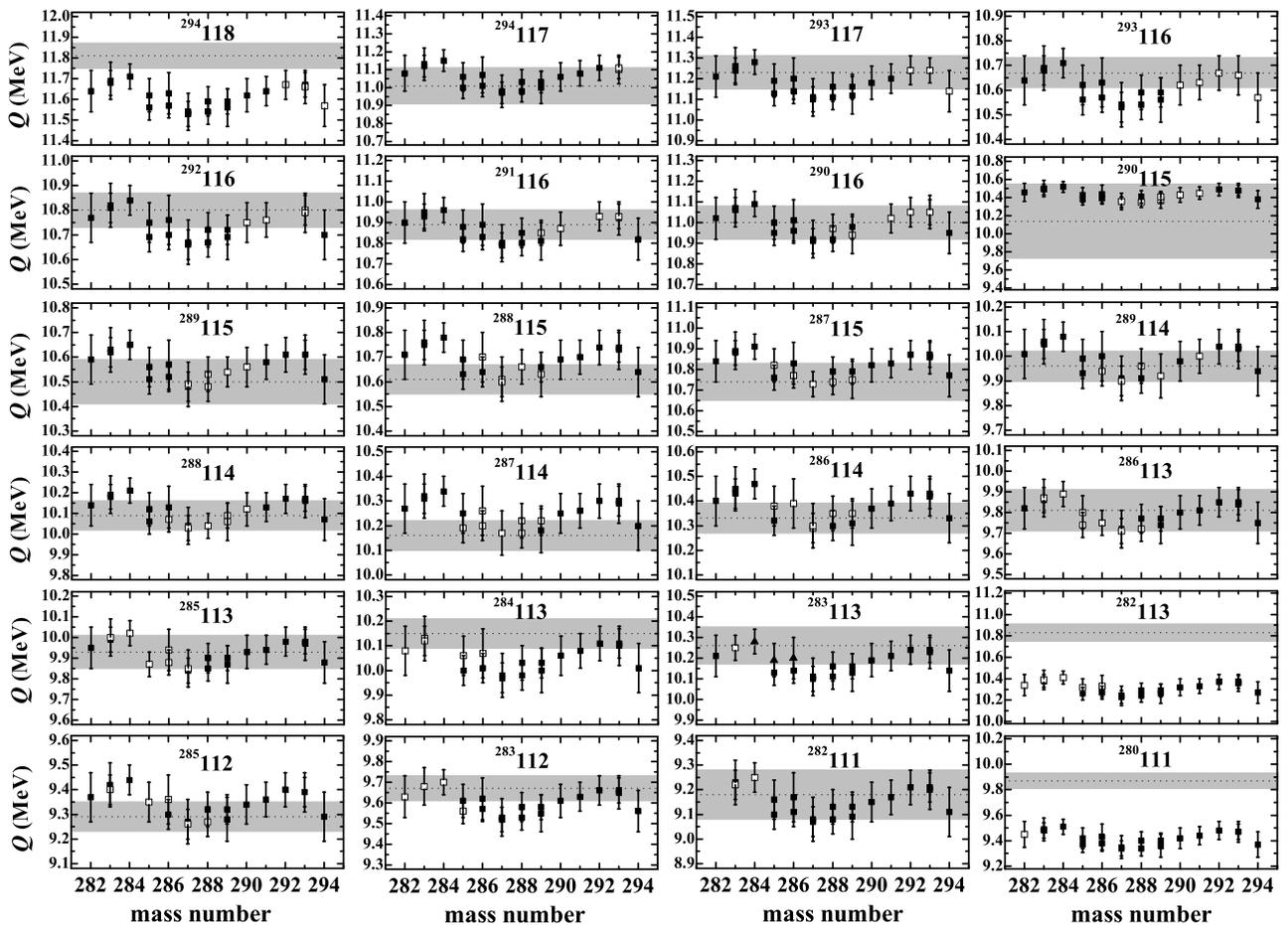}
\caption{Comparison of $\alpha$-decay $Q$ values with Eq. (\ref{EE})
(the rectangles with error bars) and experimental ones \cite{YY1}
(shaded area) of recently synthesized heaviest SHN. The horizontal
ordinate denotes the mass numbers of the reference nuclei. Since the
experimental $Q$ values of the reference nuclei include
uncertainties, the calculated ones also display error bars. The
results from Eqs. (\ref{BB}), (\ref{CC}) and (\ref{DD}) which are
special cases of Eq. (\ref{EE}), are presented separately marked by
hollow rectangles.}
\end{center}
\end{figure*}
%%%%%%%%%%%%%%%%%%%%%%%%%%%%%%%%%%%%%%%%%%%%%%%%%%%%%%%%%%%%%%%%%%%%%%%%%%%%%%%%%%

In order to test the applicability of Eq. (\ref{EE}), we compute the
$Q$ values of recently synthesized heaviest SHN with the help of
their neighbors, and the results are listed in Fig. 1 compared with
experimental ones. The results obtained with Eqs. (\ref{BB}),
(\ref{CC}) and (\ref{DD}) are marked by distinguishable symbols. For
the nuclei except $^{294}118$, $^{290}115$, $^{282}113$ and
$^{280}111$, our approach reproduces the measured values quite
accurately with a root-mean-square deviation $\sqrt{\langle \sigma
^{2}\rangle }=0.077$ MeV and an average deviation $\langle \sigma
\rangle =0.064$ MeV for central values from 380 reference-target
combinations. It is thus very practical that Eq. (\ref{EE}) can be
reliably applied to the $Q$ values of the as-yet-unobserved SHN with
the help of known nuclei which is the most effective method to the
$Q$ values at present. As three simple cases of Eq. (\ref{EE}), Eqs.
(\ref{BB}), (\ref{CC}) and (\ref{DD}) work even better with
$\sqrt{\langle \sigma ^{2}\rangle }=0.052$ MeV and $\langle \sigma
\rangle =0.043$ MeV from 96 reference-target combinations, which are
very convenient to be used and are sufficient for predictions of $Q$
values generally, though they are simple in formalism. In addition,
the agreement between the experimental and theoretical values has
additional significance. Since the $Q$ values of the reference
nuclei are taken from the experimental measurements in calculations,
the agreement suggests that the experimental data themselves are
consistent with each other, which indicates that the experimental
observations and measurements of the SHN are reliable to a great
extent. These SHN still await independent verification by other
laboratories, which is not easy because the new SHN form an isolated
island that tends to be not linked through $\alpha$-decay chains
with any known nuclei, making the theoretical supports become
important and necessary. For the nuclide $^{290}115$, the
experimental value of $10.14\pm0.41$ MeV carries a large uncertainty
while the $Q$ value is about 10.4 MeV according to Eq. (\ref{EE}),
which requires a more precise experimental measurement.

The shell closures should play a particular important role in the
superheavy system. However, modern theoretical approaches disagree
on the position of the closed shells. For instance, the
macroscopic-microscopic models with various parameterizations
predict the shell gaps at $Z = 114$ and $N = 184$ \cite{P1, AB}.
Skyrme-Hartree-Fock calculations favor $Z=124,126$ and $N=184$
\cite{SHF1,SHF2} while the relativistic mean field models favor
$Z=120$, $N=172$ \cite{SHF2,RMF,SHF3} and $Z=120$, $N=184$
\cite{RMF0}. The magic numbers $Z=132$ and $N=194$ were predicted
from the discontinuity of the volume integral at shell closures
\cite{PM}. The reason for this uncertainty lies in incomplete
knowledge of the nuclear force and the difficulty of many-body
techniques. It is well known that the shell effect on the $\alpha$
radioactivity is related to the $Q$ value. For the $\alpha$-decay of
the nuclei being not close to the shell closures, due to a parent
nucleus and its daughter nucleus sharing the same odevity of both
the proton and neutron numbers, the shell correction (also pairing
correction) energies to their masses could be canceled to a large
extent leading to a small correction to a $Q$ value compared with
the contributions of the Coulomb and symmetry energies within
semi-empirical formulas \cite{dong}, and even these small shell
energies to the $Q$ values turn out to be nearly a constant in a
local region of particle numbers which confirms that the shell
energies hardly take effect in Eq. (\ref{EE}). Most importantly, the
agreements between the estimated and experimental results in turn
show this point. Once the parent nucleus or the daughter one has
neutron and/or proton magic numbers or the shell gaps are crossed,
the $Q$ value shows an irregular behaviour. Since the shell energy
is excluded in Eq. (\ref{EE}), it should show some discrepancies for
nuclei around shell closures. Yet, it would help us to investigate
the shell structure by comparing the experimental and calculated $Q$
values. All the theoretical calculated $Q$ values of $^{294}118$,
$^{282}113$ and $^{280}111$ based on Eq. (\ref{EE}) are lower than
the experimental ones, which is possibly attributed to the likely
shell gaps at $Z=120$ for $^{294}118$, and at $N=166$ for
$^{282}113$ and $^{280}111$. In Ref. \cite{gap}, it is suggested
that $N=166$ is a neutron shell gap in a certain region within
relativistic mean field models. Yet, to confirm the existence of
shell gaps positively is not easy due to the insufficient
experimental observations. Apart from the shell effects, dramatic
shape changes could also affect the energy of $\alpha$-decay, as
pointed out in Ref. \cite{shape}. Nevertheless, the deviations are
not more than 0.5 MeV in general. However, it should be much easier
to confirm the non-existence of shell closures. Once the $Q$ values
behave quite regularly in a local range, a magic number should not
appear here. For the eight nuclides of elements 116 and 114
($^{290-293}116$ and $^{286-289}114$) together with the six nuclei
with a neutron number $N=174$ ($^{290}116$, $^{289}115$ and
$^{288}114$) and $N=172$ ($^{287}115$, $^{286}114$ and $^{285}113$),
the experimental $Q$ values can be reproduced very accurately that
confirms $Z=114$ and $N=172$ are not shell closures in the
considered region. Of course, one cannot rule out the possibility
that they appear as magic numbers in other mass regions.

%%%%%%%%%%%%%%%%%%%%%%%%%%%%%%%%%%%%%%%%%%%%%%%%%%%%%%%%%%%%%%%%%%%%%%%%%%%%%%%%%%%  Figure 2
\begin{figure}[htbp]
\begin{center}
\includegraphics[width=0.5\textwidth,clip]{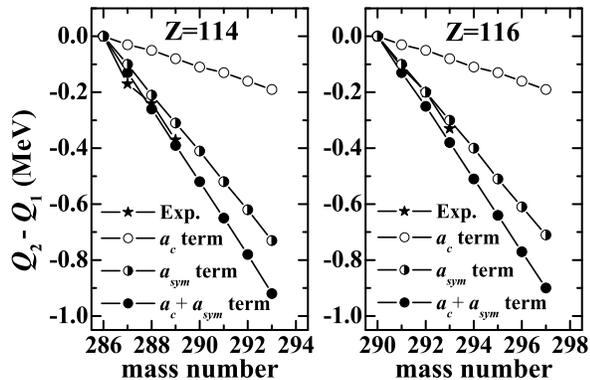}
\caption{Contributions of the $a_{sym}$ and $a_{c}$ terms to
$Q_{2}-Q_{1}$ in Eq. (\ref{CC}) taking the elements 114 and 116 as
examples ($^{286}114$ and $^{290}116$ as reference nuclei with decay
energies $Q_{1}$, respectively). The experimental data, if
available, are also shown for comparison.}
\end{center}
\end{figure}
%%%%%%%%%%%%%%%%%%%%%%%%%%%%%%%%%%%%%%%%%%%%%%%%%%%%%%%%%%%%%%%%%%%%%%%%%%%%%%%%%%%

We now turn to the effect of the symmetry energy on $\alpha$-decay.
The $a_{sym}$ term (difference of the symmetry energy effect between
the reference and target nuclei) contributes by about $45\%$ in Eq.
(\ref{BB}), $80\%$ in Eq. (\ref{CC}) and $35\%$ in Eq. (\ref{DD}) to
the $\Delta Q=Q_{2}-Q_{1}$ for the SHN in Fig. 1, suggesting its
important role in the correlations between the $Q$ values of SHN.
The large contribution of $a_{sym}$ term in Eq. (\ref{CC}) is of
particular importance as discussed below. The most significant
experimental conclusion is these observed superheavy elements
generally display a trend of increased stability with larger neutron
number, which is almost attributed to the larger symmetry energy
that lowers the $Q$ values. In order to illuminate this conclusion
more obviously, we plot in Fig. 2 the contributions of the $a_{sym}$
and $a_{c}$ terms to $Q_{2}-Q_{1}$ in Eq. (\ref{CC}). One can find
that the $a_{sym}$ term contributes much more greatly than the
$a_{c}$ term. Therefore, due to the inclusion of the effect of
symmetry energy, apart from the theoretical estimations being able
to agree with the experimental values, the $Q$ values reduce much
more rapidly as $N$ increases, and hence a superheavy element
becomes longer-lived against $\alpha$-decay with increasing $N$. In
other words, it is the symmetry energy that primarily enhances the
stability against $\alpha$-decay with larger neutron number for
these synthesized SHN not around shell closures.

We have investigated some aspects of the $\alpha$-decay of SHN. The
main conclusions are summarized as follows: (1) A simple formula for
the correlation between the $\alpha$-decay $Q$ values of the SHN has
been proposed, which works very well for an estimation of the
$\alpha$-decay energies of the recently synthesized SHN. They thus
allow us to reliably predict the $Q$ values of the still unknown SHN
with a good accuracy, and is going to be very useful for the future
experiment design. Also, the agreements between the calculated and
experimental values indicate the reliability of the experimental
observations and measurements on these synthesized SHN to a great
extent. (2) $Z=114$ and $N=172$ turn out to be not shell closures
for the presently observed superheavy region experimentally. (3) The
observed increase of $\alpha$-decay half-lives with increasing
neutron number, i.e., the increased stability of these SHN not
around shell closures with larger neutron number, is primarily
attributed to the effect of the symmetry energy.

Jianmin Dong is thankful to Prof. Jianzhong Gu, Hongfei Zhang and
Ms. Fang Zhang for providing useful help. This work is supported by
the National Natural Science Foundation of China
(10875151,10575119,10975190,10947109), the Major State Basic
Research Developing Program of China under No. 2007CB815003 and
2007CB815004, the Knowledge Innovation Project (KJCX2-EW-N01) of
Chinese Academy of Sciences, CAS/SAFEA International Partnership
Program for Creative Research Teams (CXTD-J2005-1), the Funds for
Creative Research Groups of China (Grant 11021504) and the financial
support from DFG of Germany.

%\end{CJK*}

\end{document}